\documentclass[10pt,aps,prd,twocolumn,superscriptaddress,amsfonts,amssymb,amsmath,eqsecnum,nofootinbib,floatfix]{revtex4}
\usepackage[dvips]{graphicx}
\usepackage{psfrag}
\usepackage{amsmath,amssymb,amsthm}
\usepackage{euler}
\begin{document}
\title{Corner reflectors and Quantum-Non-Demolition Measurements \\
	in gravitational wave antennae}
\author{V. B. Braginsky and S. P. Vyatchanin\\ 
    Physics Faculty, Moscow State University,
    Moscow 119992, Russia \\
    e-mail: vyat@hbar.phys.msu.ru}
\date{\today}
\begin{abstract}
We propose  Fabry-Perot cavity with corner reflectors instead of spherical
mirrors to reduce the contribution of thermoelastic noise in the coating which
is relatively large for spherical mirrors and which prevents the sensitivity
better than Standard Quantum Limit (SQL) from being achieved  in laser
gravitational wave antenna. We demonstrate that thermo-refractive noise in
corner reflector (CR) is substantially smaller than SQL.  We show that the
distortion of main mode of  cavity with CR caused by tilt and displacement of
one reflector is smaller than for cavity with spherical mirrors. We also
consider the distortion caused by small nonperpendicularity of corner facets
and by optical inhomogeneity of fused silica which is proposed as a material
for corner reflectors.

\end{abstract}

\maketitle

\section{Introduction}

The existing to-day's multi-layer dielectric coating on optical mirrors
allows to realize very high resolution experiments
(see e.g. \cite{kimbleA}). The reflectivity  $R$ in the best
optical coating has reached the level of $(1-R)\simeq 10^{-6}$
\cite{kimbleA,kimble,rempe}
(commercially available $(1-R)\simeq 10^{-5}$), and there are many reasons
to expect that further improvement of coating technologies will permit
to obtain the value of $(1-R)\simeq 10^{-9}$. With the value of
$(1-R)\simeq 10^{-5}$ it is possible to realize the ring down time
$\tau_{FP}^*\simeq 1$~sec in $4$ km long Fabry-Perot (FP) resonators
which are the basic elements in laser interferometer gravitational wave
antennae (project LIGO \cite{abr1,abr2}). This relatively large value
of $\tau^*_{FP}$  permits to have relatively small value of the ratio of
$\sqrt{\tau_{av}/\tau_{FP}^*}\simeq 7\times 10^{-2}$ (if the averaging
time $\tau_{av}\simeq 5\times 10^{-3}$~sec). This ratio is the limit
for the squeezing factor which may be obtained if QND procedure of
measurement in such FP resonator is used \cite{bh1,bh2}. Such a
procedure will allow to circumvent the Standard Quantum Limit
(SQL) of sensitivity (see details in \cite{f1}).

Few years ago the role of the thermoelastic noise in the bulk of the mirrors
was analyzed \cite{bgv}. This analysis has shown that if the laser beam spot
size on the mirror surface is sufficiently large then the small value of the
thermal expansion coefficient $\alpha_{SiO_2}\simeq 5\times 10^{-7}\ K^{-1}$
of fused silica will permit to circumvent the SQL sensitivity by the factor of
$\simeq 0.1$ (if the thermoelastic noise is the only source of noises). The
consequent analysis of thermoelastic noise in the coating itself unfortunately
predicts that the limit of sensitivity will be close to the SQL of sensitivity
\cite{bv,fejer,fejer2,03a1BrSa}. The origin of this obstacle is relatively big
numerical value of thermal expansion coefficient $\alpha_{Ta_2O_5}\simeq
5\times 10^{-6}\ K^{-1}$ of amorphous $Ta_2O_5$ \cite{03a1BrSa} which is used
in the best coatings as well as relatively big number of layers (usually
20-40) which is necessary to have small value of $(1-R)$. For LIGO project
these limitations may be illustrated by the following numerical values. The
SQL sensitivity of detectable amplitude of the perturbation of the metric is
equal to \cite{thorne}

\begin{align}
\label{ShSQL}
\sqrt{S_{h}^{SQL}(\omega)} & =\sqrt{\frac{8\hbar}{m \omega^2 L^2}}
        \simeq  2\times 10^{-24}\ \mbox{Hz}^{-1/2},
\end{align}

where $m=40$~kg is mass of test mass, $L=4$~km is distance between them,
$\omega=2\pi\times 100$~s$^{-1}$ is
observation frequency. Here and below the estimates are calculated for
numerical parameters listed in Appendix \ref{param}.
At the same time, according to the measurement \cite{03a1BrSa},
 the limit of sensitivity of such an antenna only due to the
thermoelastic noise in the multi-layer $Ta_2O_5+SiO_2$ coating on $SiO_2$
substrate has to be between \cite{bv}
\begin{align}
\label{TDcoat}
\sqrt{S_h^{TD\, coat}(\omega)}\simeq \big(0.6 \div1.4\big) \times 10^{-24}\
	\mbox{Hz}^{-1/2}
\end{align}

\begin{figure}
\psfrag{a}{\bf a)}
\includegraphics[width=0.5\textwidth]{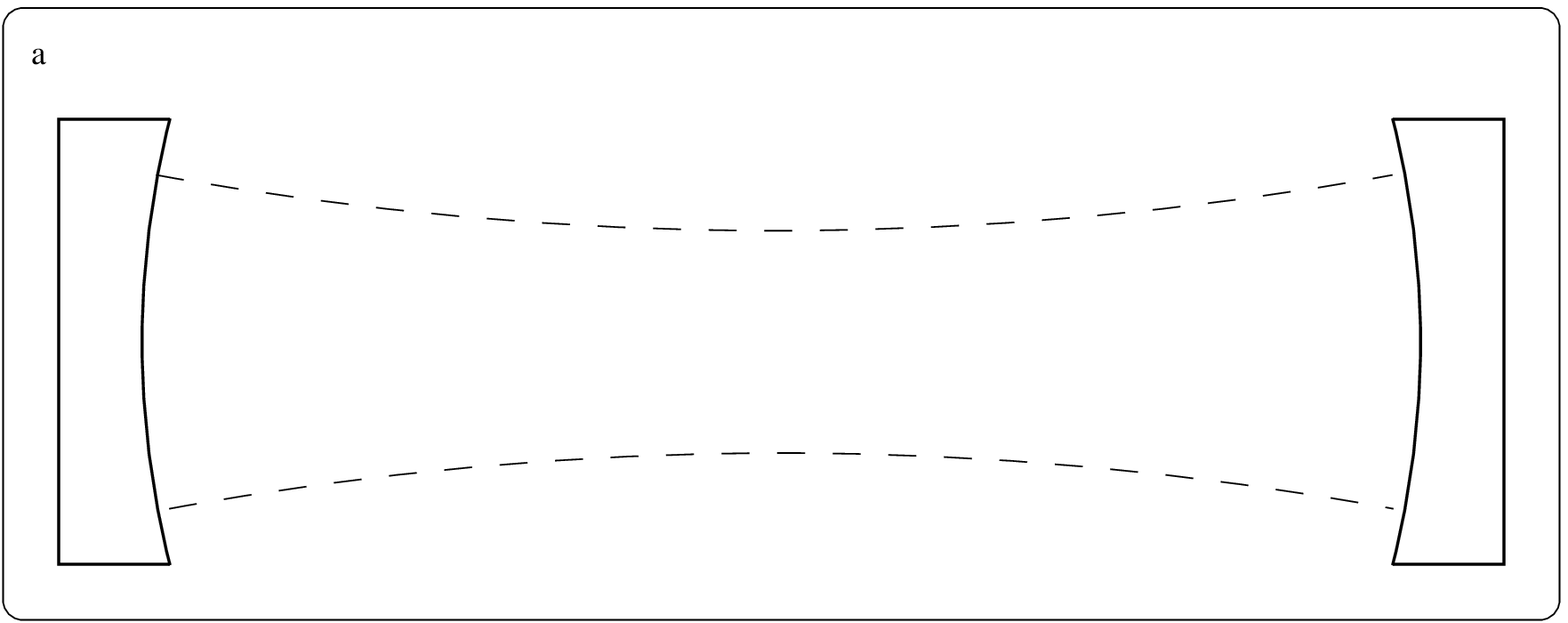}
\psfrag{b}{\bf b)}
\psfrag{e}{$\frac{\pi}{2} +\epsilon$}
\includegraphics[width=0.5\textwidth]{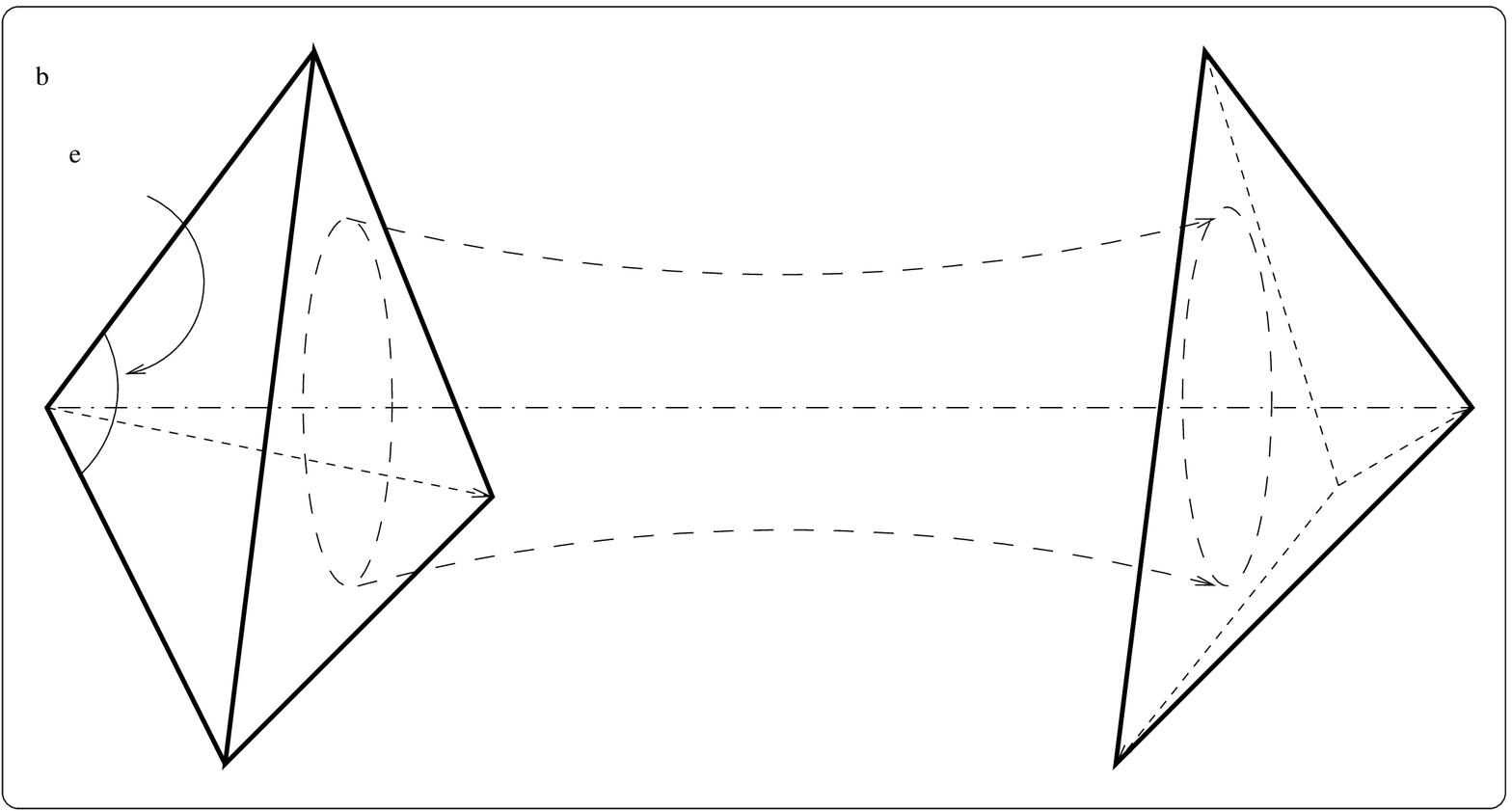}
\psfrag{c}{\bf c)}
\psfrag{e}{$\hspace{-2mm}\frac{\pi}{2}+\epsilon$}
\psfrag{K}{$ \theta$}
\psfrag{k}{$\varkappa$}
\includegraphics[width=0.5\textwidth]{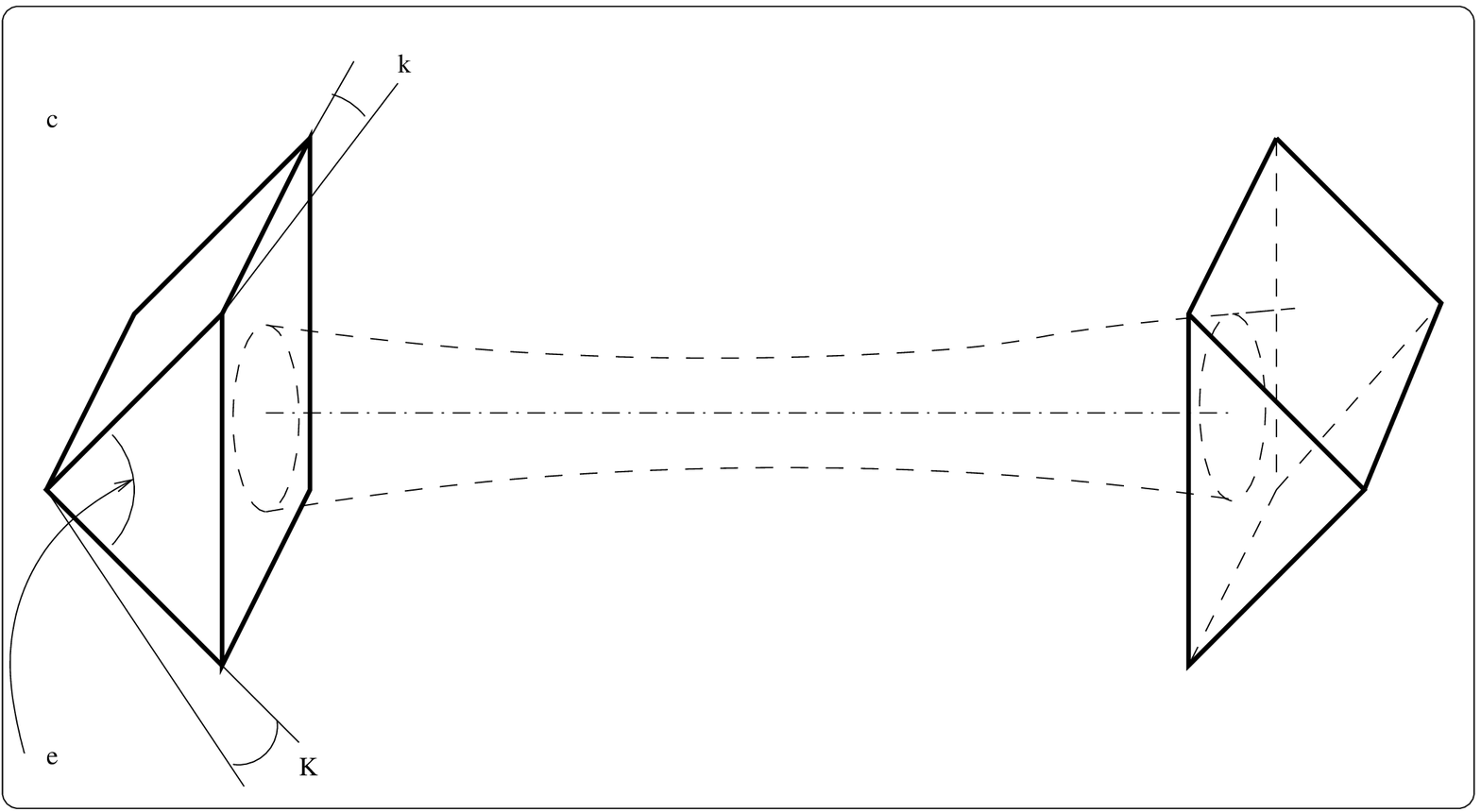}
\caption{We propose to replace  mirrors with finite value of surface curvature
(a) by 3 facets "triprism" type CR (b) or 2 facets "roof" type
CR  (c).}\label{fig1}
\end{figure}

The goal of this article is to present the analysis of another version of
optical FP cavity where the "contribution" of the thermoelastic
noise in coating is substantially reduced. The key idea of this version is
based on concept of corner optical reflector (tri-hedral or two-hedral prism).
These types of reflectors
were well known among the jewelers at least from 16-th century
(see e.g. autobiography by Benvenutto Cellini \cite{chellini}). In the 70-s
of the previous century corner reflectors (CR) installed on the Moon
allowed to test the principle of equivalence for the gravitational
defect of mass by laser ranging \cite{moon}. Here we propose to substitute
mirrors with finite value of surface curvature (see fig.\ref{fig1}~a)
by 3 facets (fig.\ref{fig1}~b) or 2 facets (see fig.\ref{fig1}~c)
corner reflectors (CR) manufactured from fused silica.
For the same radius  $R_b$ of laser beam the mass of CR is about
the same value as cylindrical mirrors (with height about equal to
radius of cylinder) especially if idle parts of CR is removed. For example for
2 facets CR the size of foot surface has to be about $45\times 45$~cm$^2$ with
total test mass about $40$~kg (the same as planned in advanced LIGO).

In the proposed scheme the stability of optical mode is provided by surfaces
lens shaping of each CR foot (as shown in fig.\ref{ideal}a).
For reflectors  manufactured from fused silica the total internal reflection
inside the reflectors is possible because refraction index $n_{SiO_2}=1.45$ is
large enough (due to Snellius law): $n_{SiO_2}\sqrt {2/3}>1 $ for 3 facets
reflector or $n_{SiO_2}\sqrt {1/2}>1 $ for 2 facets reflector.

In section \ref{sec:ideal} we consider the modes of ideal cavity
(fig.\ref{ideal}) and the distortion of the main mode structure caused by small
perturbations of different kinds: tilt angle $\theta$ (fig.~\ref{tilt}),
displacement $\delta x$ of one reflector (fig.~\ref{disp2}), expose angle
$\epsilon$ (fig.\ref{expos}).

In section \ref{comp} we compare these perturbations for cavities with
spherical mirrors and with corner reflectors and give numerical estimates for
the particular case of  laser beam radius $R_b\simeq 6$~cm (intensity of beam
decreases as $e^{-1}$ at distance $R_b$ from center) which is planned for
Advanced LIGO. Distortions of mode are undesirable in high accuracy
spectroscopic measurements because they may produce additional noise. For
example, in laser  gravitational wave  antenna the light beams from two
independently perturbed FP cavities (placed in each arm of Michelson
interferometer) will not produce completely zero field at the dark port after
the beam splitter. It is equivalent to additional noise at the dark port.

In section \ref{losses} we consider the different sources of optical losses of
CR and show that they can be at level $(1-R)\simeq 10^{-5}$ .

In cavity with CRs   it is necessary to use nevertheless relatively thin
anti-reflective coatings ($2$ -- $4$ layers) on lense shape foot. It has to be
done to keep the value $(1-R)$ at the level $\simeq 10^{-5}$.  Because this
coating is substabtialy thiner than typical  high reflective one ($20$ -- $40$
layers) used in curved mirrors, thermoelastic noise may be depressed by the
factor $\sim 10$ i.e. about one order less than SQL (the estimate
(\ref{TDcoat}) is given for $38$ layers). The "fee" for use of CR is the
additional thermo-refractive noise  \cite{bgv2} (fluctuations of temperature
produce the fluctuations of refractive index) because light beam is traveling
inside the corner reflector.  However we will show in section \ref{sec1} that
it is several times smaller than SQL for reflectors manufactured from fused
silica. Important that this thermo-refractive noise  rapidly decreases with
increasing of radius $R_b$ of beam spot as $\sim R_b^{-2}$.

In cavity with CR the thermoelastic noise in the facets
remains, but as mentioned above if the reflectors are manufactured
from fused silica and the
beam spot is large enough (see e.g. \cite{bv}) then it is possible to
circumvent SQL.

\section{The Distortions of main mode in
		FP cavity with  CR}\label{sec:ideal}

Firstly, we consider  FP cavity with  two identical
perfect  corner cube reflectors with three reflecting
facets (see fig.\ref{fig1}b):
{i)} the corner angles between the facets are exactly equal to $\pi/2$;
{ii)} the top points of the reflectors are located exactly
	on common optical axis;
{iii)} the "foots" of the reflectors have slight curvatures (shape of a lens
surface as shown in fig.~\ref{ideal}a) and they are perpendicular to the axis.
Then we consider the distortion of mode in this cavity caused by
different perturbations.

\subsection{FP resonator with perfect CR}

\begin{figure}
\psfrag{f}{\bf a)}
\psfrag{g}{\bf b)}
\psfrag{A}{$A$}
\psfrag{A'}{$A'$}
\psfrag{B}{$B$}
\psfrag{B'}{$B'$}
\psfrag{C}{$C$}
\psfrag{C'}{$C'$}
\psfrag{x}{$ x$}
\psfrag{y}{$ y$}
\psfrag{z}{$z$}
\includegraphics[width=0.5\textwidth]{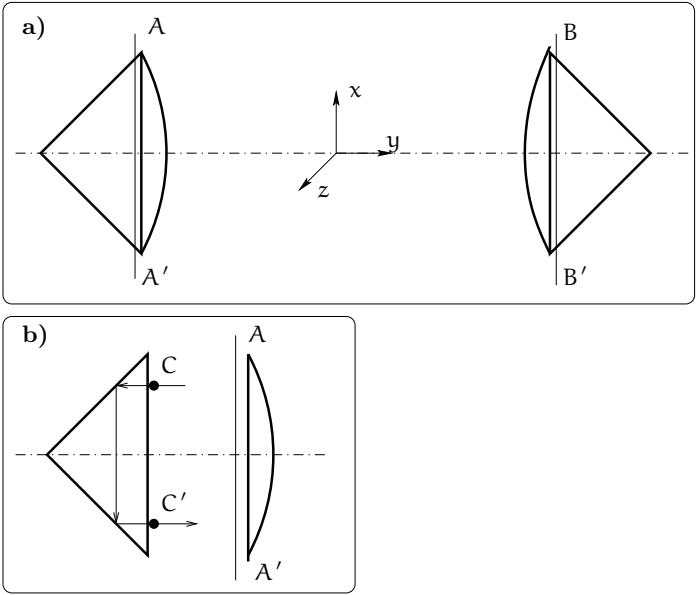}
\caption{a). The  perfect alignment of  CR assembling cavity.
	b). Each reflector can be regarded as reflector with plane
	"foot" plus a lens.}\label{ideal}
\end{figure}

We can consider that each CR consists of reflector with plane foot
surface together with spherical lens as shown in fig.\ref{ideal}b. The CR
produces mirror transformation, i.e. light beam which enters the
reflector in point $C$ is transformed into the beam leaving the reflector in point $C'$.
Using Fresnel integral one can obtain the integral equations for calculations
of eigenmode distribution:
\begin{align}
\label{ideal1}
e^{ikL}\int G_0(\vec r_1, \vec r_2)\,\Phi_2(\vec r_2)\, d\vec r_2 &=
     \lambda\tilde \Phi_1(\vec r_1),\\
\label{ideal2}
e^{ikL}\int G_0(\vec r_1, \vec r_2)\,\Phi_1(\vec r_2)\, d\vec r_1 &=
    \lambda\tilde \Phi_2(\vec r_2),\\
d\vec r_1= dx_1\, dy_1, &
    \quad d\vec r_2= dx_2\, dy_2.\nonumber
\end{align}
Here functions $\Phi_1(\vec r_1)$ and $\Phi_2(\vec r_2)$ describe distribution
of complex field amplitude emitted from imagine flat foot surfaces  of reflector 1
(left) and reflector 2 (right) correspondingly just under lenses
in planes $AA'$ and $BB'$. $L$ is the optical path between reflectors
(including path inside the reflector).  Evidently phase fronts coincide with planes $AA'$ and $BB'$ so that phases of functions $\Phi_1$ and $\Phi_2$ are constants. The notation $\tilde\Phi_1(\vec r_1)$
means the "mirror" transformation (produced by 3 facets CR)
relative to the optical axis\footnote{For 2 facets
reflector we have mirror transformation only relatively $x$ coordinate:
$\tilde\Phi_1(x, y))=\Phi_1(-x,y)$}:
$$
\tilde\Phi_1(\vec r_1)=\Phi_1(-\vec r_1),\qquad
    \tilde\Phi_2(\vec r_1)=\tilde\Phi_2(-\vec r_1)
$$
The kernel $G^0$  is:
\begin{align}
\label{G0a}
G_0(\vec r_1, \vec r_2) &= -\frac{i}{2\pi}
    e^{i\left(\frac{(\vec r_1-\vec r_2)^2}{2} -
         h_1(\vec r_1) -h_2(\vec r_2)\right)},\\
\label{hh}
h_1&=\frac{r_1^2}{r_h^2},\qquad h_2=\frac{r_2^2}{r_h^2}
\end{align}

Here we use the dimensionless transversal coordinates
$\vec r_1$ and $\vec r_2$ (at  planes $AA'$ and $BB'$)
which can be expressed in terms of physical coordinates $\vec R_1$ and $\vec R_2$
as
$$
\vec r_1=\frac{ \vec R_1}{b}, \qquad
    \vec r_2=\frac{ \vec R_2}{b}, \qquad
    b=\sqrt\frac{L}{k},
$$
where $k$ is wave vector, $h_1$ and $h_2$ are additional phase shifts produced
by spherical lenses at each reflector foot. It is easy to see that for
spherical lenses the set of eigenmodes $\Phi_1^{mn},\ \Phi_2^{mn}$ and their
eigenvalues $\lambda_{mn}$ (below we assume $\lambda_{00}=1$)
of our FP cavity are described by
generalized Gauss-Hermite functions:
\begin{align}
\Phi_{1}^{mn} &= \phi_m(x)\phi_n(y),\\
\Phi_{2}^{mn} &= (-1)^{-m+n}\Phi_{1}^{mn}=\tilde\Phi_{1}^{mn}  \\
\phi_m(x) &= \frac{1}{\sqrt{r_L}\, \sqrt{\sqrt
\pi \, 2^{m }\, m!} }\,
	H_m\left(\frac { x}{r_L}\right)\times\\
&\qquad  \times       \exp\left[-i(m+1/2)\psi-
		\frac{x^2}{2r_L^2}
	\right],\nonumber \\
\label{psi}
\lambda_{mn}&=e^{2i(m+n)\psi} ,\qquad
        \psi = \arctan\left(\frac{1}{2r_0^2}\right), \nonumber\\
h_1(r)&=h_2(r) = \frac {r^2}{2r_h^2},\quad 2r_L^2=2r_0^2+\frac{1}{2r_0^2}\\
2r_h^2 &= (2r_0^2)^2+1=2r_L^2\, 2r_0^2.
\end{align}
Here $H_m(t)$ is the Hermite polynomial of the order  $m$,
$r_0$ and $r_L$ are the radii of beam in the waist and at the
lens  correspondingly.
It is useful to write down the expressions for $r_0,\ r_L,\ r_h$ using
$g$-parameter \cite{siegman}
($R^*$ --- is the radius of wave front curvature (in cm) just after the
propagation of beam through the lens outside of the reflector):
\begin{align}
\label{g1}
g &= 1 -\frac {L}{R^*},\quad
      r_0^2 = \frac{1}{2} \,\sqrt{\frac{1+g}{1-g} }, \\
r_L^2&=\frac{R_b^2}{b^2}=\frac{1}{\sqrt{1-g^2} },\quad
      r_h^2 = \frac{R^*}{L}= \frac{1}{1-g},\\
\label{g3}
\sin 2\psi &=g,\quad \cos 2\psi= \sqrt{1-g^2}.
\end{align}

We are interested in the main mode $\Phi_{1}^{00}(x,y)$ of resonator
(amplitude distributions of left and right
reflectors obviously coincide  with each other for the main mode:
$\Phi_{1}^{00}(x,y)=\Phi_{2}^{00}(x,y)$).

\subsection{Distortion due to the Tilt of CR}\label{sec:tilt}

\begin{figure}
\psfrag{f}{}
\psfrag{a}{\bf a)}
\psfrag{b}{\bf b)}
\psfrag{d}{$l$}
\psfrag{t}{$\theta$)}
\psfrag{x}{$ x$}
\psfrag{z}{$z$}
\psfrag{de}{$\delta x\simeq l\, \theta$}
\includegraphics[width=0.5\textwidth]{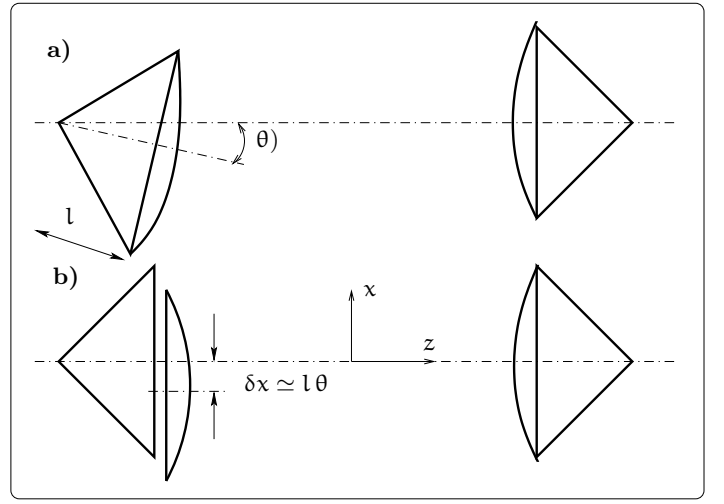}
\caption{a). Small tilt of the left CR around its head.
	b) This tilt  is equivalent to untilted reflector and displaced
	lens.}\label{tilt}
\end{figure}

Here we consider the  main mode $\breve \Phi_{1}^{00}(x,y)$ perturbed due to
tilt misalignment shown in fig.~\ref{tilt}a. We expand the perturbed main mode
into series over the set of unperturbed modes limiting ourselves to the lowest
(dipole) approximation:
\begin{align}
\label{pert1}
\breve \Phi_{1}^{00}(x_1,y_1)&\simeq \Phi_{1}^{00}(x_1,y_1) -
	\alpha_1^{\rm tilt} \Phi_{1}^{10}(x_1,y_1),\\
\label{pert2}
\breve \Phi_{2}^{00}(x_2,y_2)&\simeq \Phi_{2}^{00}(x_2,y_2) +
	\beta_1^{\rm tilt} \Phi_{2}^{10}(x_2,y_2).
\end{align}

The tilt of CR around its head through a small angle of $\theta$
can be considered as untilted reflector with
lens displaced a small distance $\delta x\simeq l\theta$
perpendicular to optical axis ($l$ is the dimensionless distance
between the foot and head of CR, it is illustrated in
fig.~\ref{tilt}b). For this case the perturbations of the main mode  can be
described by dipole coefficients defined in (\ref{pert1}, \ref{pert2}):
\begin{align}
\label{atilt1}
\alpha_1^{\rm tilt} &\simeq
	l\theta\, \frac{ g\,(1-g)}
	{\sqrt 2\,(1-g^2)^{3/.4}},\\
\label{btilt1}
\beta_1^{\rm tilt} &\simeq 	l\theta\,
	\frac{(1-g)\,\big[\cot \big(\psi\big) -i\big]}
	{2\sqrt 2 \, (1-g^2)^{1/4} },\\
\label{btilt1b}
|\beta_1^{\rm tilt}| &\simeq
	 \frac{l \theta}{2}\,\sqrt[4]{\frac{1-g}{1+g}}
\end{align}
See details of calculations in Appendix \ref{Atilt}.

Note that distortion produced by tilt of two facets CR around perpendicular
axes (angle $\varkappa$ on fig.~\ref{fig1}c) can be described by the same
formulas as tilt of spherical mirror.

\subsection{The Distortion due to the
        Displacement of CR}\label{sec:disp}

\begin{figure}
\psfrag{dd}{}
\psfrag{d}{$\delta x$}
\psfrag{x}{$ x$}
\psfrag{y}{$ y$}
\psfrag{z}{$z$}
\includegraphics[width=0.5\textwidth]{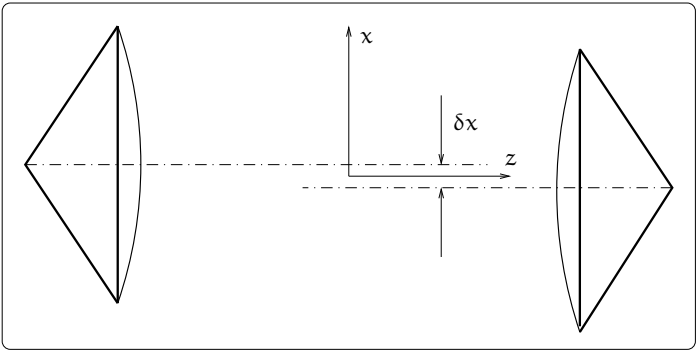}
\caption{One CR is displaced
	relative to another one.}\label{disp2}
\end{figure}

One CR can be displaced by a small distance $\delta x$ so that
optical axes of reflectors do not coincide with each other as it is shown in
fig.~\ref{disp2}. For this case the perturbations of the main mode  can be
described by dipole coefficients defined by formulae (\ref{pert1},
\ref{pert2}). Denoting the dipole coefficients as $\alpha_1^{\rm displ}$ and
$\beta_1^{\rm displ}$ one can obtain:
\begin{align}
\label{alphadisp}
\alpha_1^{\rm displ} &=\beta_1^{\rm displ} \simeq
	 \frac{-
         \delta x  }{2\,\sqrt 2\,  }
	\left(\frac{-2ig}{\sqrt[4]{1-g^2}}  + \sqrt[4]{1-g^2}\right)
\end{align}
See details of calculations in Appendix~\ref{Adisp}.

\subsection{The Distortion of Expose Angle}\label{sec:expos}

\begin{figure}
\psfrag{a}{\bf a)}
\psfrag{b}{\bf b)}
\psfrag{c}{\bf c)}
\psfrag{p2}{$ \frac{\pi}{2}+\epsilon$}
\psfrag{g}{$ \gamma$}
\includegraphics[width=0.5\textwidth]{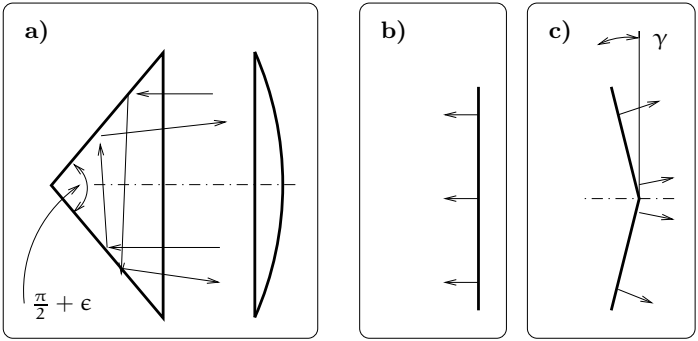}
\caption{(a). Expose perturbation: the angle between facets of corner
        reflector differs from
	direct angle by a small value of $\epsilon$.
	It produces the  transformation of incident
	plane wave front (b) into a "broken" front of reflected
	wave (c).}\label{expos}
\end{figure}
Here we consider the case when, for example the left reflector (2-hedral prism
shown in fig.~\ref{fig1}c) has a non-perfect perpendicular facets, so that
expose angle between them differs from $\pi/2$ by a small angle of $\epsilon$
(fig.~\ref{expos}a). Then the plane front of incident wave after reflection
from the reflector is transformed into a broken surface consisting of two plane parts
declined to the incident wave front by an angle of $\gamma =2\epsilon$ as shown in
fig.~\ref{expos}b,c.

This statement is also correct for tri-hedral reflectors
(shown in fig.~\ref{fig1}b) with the exception of numerical factor:
if only one facet is declined by angle of $\epsilon$ from the normal position
(and other two facets are non-perturbed)
the angle $\gamma$ will be equal to $\gamma=2\epsilon\sqrt{2/3}$.

Again we  can expand the perturbed main mode over the set of unperturbed
modes of ideal cavity keeping only the lowest first-order non-vanishing term of expansion:
\begin{align}
\breve \Phi_{1}^{00}(x_1,y_1)&\simeq \Phi_{1}^{00}(x_1,y_1) +
	\alpha_2^{\rm expose} \Phi_{1}^{20}(x_1,y_1),\\
\breve \Phi_{2}^{00}(x_2,y_2)&\simeq \Phi_{2}^{00}(x_2,y_2) +
	\beta_2^{\rm expose} \Phi_{2}^{20}(x_2,y_2)
\end{align}
(due to the symmetry of this kind perturbation the  dipole term is null).
Calculation gives the following value for $\alpha_2$:
\begin{align}
\label{expos1}
\alpha_2^{\rm expose} &\simeq
	\frac{i\gamma L}{4\sqrt 2 \, \pi\, b\sqrt[4]{1-g^2}}\times
	\frac{\big(g+i \sqrt{1-g^2}\big)^2}{i\, g\, \sqrt{1-g^2}}\\
\label{expos2}
|\alpha_2^{\rm expose}| &= |\beta_2^{\rm expose} |\simeq
	\frac{L\gamma }{4\sqrt{2}\, \pi\, b\,g(1-g^2)^{3/4}}
\end{align}
See details in Appendix~\ref{Aexpos}.

\section{Comparison of FP cavities with CR and
        with spherical mirrors}\label{comp}

Recall that  uncontrollable perturbations of the mode produce additional noise: in
laser interferometer gravitational antenna the signal at dark port will
contain  additional noise with power proportional to the square of distortion
coefficients $\sim|\alpha|^2, |\beta|^2$. In this section we compare
numerically the
distortion of the main mode in traditional FP cavity with spherical mirrors (SM
cavity) with FP cavity assembled by CR (CR cavity).

The distortion of the main mode in SM cavity caused by  small displacement and
tilt of one mirror  can be also described by coefficients $\alpha_1$ of
expansion (\ref{pert1})\cite{our, our2}:
\begin{align}
\label{tiltsph}
\alpha_1^{\rm tilt,\ Sph}&=
	\frac{1}{\sqrt 2 \, (1-g^2)^{3/4}}\left(\frac{\theta \, L}{b}\right),\\
\alpha_1^{\rm displ,\ Sph}&=
	\left(\frac{(1-g)^{1/4}}{\sqrt{2}(1+g)^{3/4}}\right)\delta x
\end{align}
For estimates for both SM and CR cavity we use the parameters of cavity
proposed for Advanced LIGO \cite{our}:
\begin{align}
r_L &= 6/2.6\simeq 2.3,\quad g=0.982. \quad
\end{align}
These parameters correspond to the radius of laser beam $R_b$
at the reflector surface of about $R_b\simeq 6$~cm.
Assuming additionally that dimension length $l^*$  from foot to top is
equal to $l^*=20$~cm, i.e.
$$l=\frac{20\, {\rm cm}}{ b}\simeq 7.7\, ,
$$
and dimension displacement $\delta x^*=b\,\delta x$
we obtain the following estimates for SM cavity:
\begin{align}
\label{estsph}
\alpha_1^{\rm tilt,\ SM}&=
	0.013\left(\frac{\theta }{10^{-8}}\right),\\
\alpha_1^{\rm disp,\ SM}&=0.0059
        \left(\frac{\delta x^* }{0.1\ \mbox{cm}}\right)
\end{align}
and  for CR cavity:
\begin{align}
\label{estcr}
\alpha_1^{\rm tilt,\ CR}&=
	1.2\times 10^{-7}\left(\frac{\theta }{10^{-8}}\right),\\
\label{estcr2}
\alpha_1^{\rm displ,\ CR}&=0.06 \left(\frac{\delta x^* }{0.1\ \mbox{cm}}\right),\\
\label{a1e}
\alpha_2^{\rm expose}&=0.11 \left(\frac{\gamma }{10^{-6}}\right).
\end{align}

We see that CR cavity  is substantially more stable to tilt and less stable to
displacement than SM cavity. However, the total requirements for SM cavity looks
more tough if one compares the estimates (\ref{estsph} and \ref{estcr2}).
Indeed to keep control of tilt in SM cavity  with accuracy  $\theta\simeq
10^{-8}$~rad (see  (\ref{estsph})) one has to operate the positioning  system
with  accuracy about $ l b\, \theta\simeq 2\times 10^{-6}$~cm. But the same
level of displacement distortion in CR cavity (see (\ref{estcr2})) one can
obtain by displacement control with much lower accuracy:  $\le 1$~mm only (!).

\paragraph*{The requirement for an expose angle in CR cavity} looks also
acceptable: for accuracy of manufacturing $\epsilon\simeq 3\times 10^{-7}$
(commercially available prisms have $\epsilon\simeq \pm 1\times 10^{-5}$)
and hence $\gamma =2\epsilon \sqrt{2/3}\simeq 4.9\times 10^{-7}$
we have to  take into account that three angles between facets (increasing
factor $\sqrt 3$)
in each of two tri-hedral reflector (one more increasing
factor $\sqrt 2$) may be independently perturbed (so the total
increasing factor is equal to $\sqrt 3 \times \sqrt 2=\sqrt 6$):
$$
\sqrt{\sum\big(\alpha_{2,\, i}^{\rm expose}\big)^2}
	\simeq \sqrt {6}\, \alpha_2^{\rm expose}
	\simeq 0.13
$$

\paragraph*{Optical inhomogeneity} is one more source of perturbation which is
specific to CR: the refraction index of fused silica (which
CR is manufactured from) changes over the value of
$\delta n\simeq 2\times 10^{-7}$ along the length $\Delta l\simeq 10$~cm
\cite{gary}.  To estimate negative influence of
this effect we can consider the model task using the fact that distance scale $\Delta
l$ of refraction index perturbation
is about the dimension $b l$ of the reflector. Let one half of corner
reflector has a perturbed refraction index $n+ \Delta n(x)$ which depends on
a transversal coordinate $x$ only:
\begin{equation}
\Delta n(x) =\left\{
        \begin{array}{lll}
        \Delta n &=-\delta n\left(1-\frac{x}{l}\right)& \mbox{if}\quad x>0,\\
        \Delta n &=0 & \mbox{if}\quad x\le 0
        \end{array}
        \right.
\end{equation}
In this case the beam after reflection from such a reflector will have a
broken wave front as shown in fig.~\ref{expos}c with angle
\begin{equation}
\label{inhomo}
\gamma = \frac{\delta n}{4n}
\end{equation}
Hence one can estimate the value of perturbation due to inhomogeneity using
formula (\ref{expos2}). It is obvious that for perturbation with another
dependence of $\Delta n$ on space coordinates the formula (\ref{inhomo}) must
change this estimate  by the factor of about unity or slightly larger. So for
estimates we use the $4$ times greater
value of $\gamma$  than (\ref{inhomo}). In that case for two
reflector with independently perturbed refraction index (factor $\sqrt 2$) we
obtain:
\begin{align} \gamma =\frac{\sqrt 2 \, \delta n}{n}&\simeq 2\times
10^{-7},\quad \alpha_2^{\rm inhomo}\simeq 0.011 \end{align}

The two last kinds of  perturbations (expose angle and inhomogeneity) depend
only on manufacturing procedure and there is a hope they can be decreased due
to the improvement of manufacturing culture.

\section{The optical losses}\label{losses}

The loss coefficient for CR cavity must be  small --- about 10~ppm. We
consider the following sources of losses.

\paragraph*{Fundamental losses on edge}
 are produced by diffraction on edge where two facets meet.
Qualitatively it can be described as two surface waves outside CR (bounded
with waves inside due to complete internal reflection) meet at edge producing
diffractional scattering (we acknowledge to
F.Ya.Fhalili pointed out the existtence of this kind losses).

To our best knowledge nobody performed rigorous analysis of this problem.
So we propose the consideration to estimate this effect: (a) using
the formulas for complex coefficient of reflection for plane wave from plane
infinite boundary between two media for the case of internal reflection
(see e.g. \cite{ll}) one can construct the solution {\em inside} CR;
(b) using the boundary condition one can obtain the fields along {\em outside}
surface of CR;
(c) applying Green's formula one can calculate the radiation field in far
wave zone and total diffractional power. The most vulnerable for critics
item of this consideration is (a) --- applying the formulas for infinite
boundary to corner configuration.

We apply this consideration for the case of incident wave polarized along
edge of CR of "roof" type (fig.~\ref{fig1}c). For this particular case (with
obvious assumption that magnetic permittivity $\mu=1$) the
all field components of constructed solution are smooth inside CR and on
outside surface CR. Our calculation gives the following loss coefficient:
\begin{equation}
\label{funddiff}
(1-R)_{d}\simeq \frac{0.4 \lambda}{R_b}\simeq 0.7\times 10^{-5}
\end{equation}
where $\lambda$ is the optical wavelength,
(see details in Appendix \ref{Adiff}).

Note that our  consideration of incident wave polarized
perpendicular to the edge of  CR allows to construct smooth solution inside CR but
this solution on the outside surface will have break of components of
electrical fields at the edge.
Thus to get a reliable confirmation of the approximatetive estimate
(\ref{funddiff}) it is necessary either to find a rigorous analytical
solution of this problem or to perform straightforward numerical calculation.

\paragraph*{The losses on non-perfect edge}
 is produced by scattering of the plane optical wave on
the non-perfect "ridges" where two facets meet. The edge of facets intersection
with uncontrollable width of $\Delta s\le 0.5\ \mu$m will produce optical losses
which can be roughly estimated as following:
$$
(1-R)_{\rm non-perfect}\le \frac{\Delta s}{R_b}\simeq 0.8\times 10^{-5}
$$
In other words it satisfies our initial condition to obtain $\tau^*_{\rm
FP}\simeq 1$s.

\paragraph*{Optical losses of material.}
Internal optical losses in purified fused silica at the level
$\gamma_{\rm loss}\simeq 0.5$~ppm/cm \cite{ligo2}
give the loss coefficient about
$$
(1-R)_{opt.loss}\simeq \gamma_{\rm loss}2bl\simeq 0.8\times 10^{-5}
$$

\paragraph*{Losses in anti reflective coating. }
As we mentioned in Introduction
it will be necessary to use  anti reflective coating on the
bottom surfaces of CR. Calculations which we omit here shows that to keep
the value of $(1-R)$ at the level $\sim 10^{-5}$ it is sufficient to use
2-4 anti reflective layers of coating.

\section{Thermo-refractive noise }\label{sec1}


The origin of thermo-refractive noise is thermodynamic (TD) fluctuations of
temperature which produce fluctuations of phase of light traveling inside
the CR through dependence of refractive index $n$ on
temperature $T$: $\beta=dn/dT\ne 0$ \cite{bgv2}. One can estimate TD
temperature fluctuations using the model of infinite layer with width $l_c$
($0\le z\le l_c$). We additionally assume
that layer is in vacuum and its both surfaces are thermally isolated (thermal
radiation in accordance with Stefan-Boltzmann law is so small that this assumption
is quite correct). If light (Gaussian beam)
with radius $R_b$
travels through the layer perpendicular to its surface the fluctuations
$\varphi$ of light phase during time $\tau$
will be defined by TD temperature fluctuations $u$
averaged over the cylinder  $\pi R_b^2 l_c$:
\begin{equation}
\label{phi}
\varphi = kl_c\,\beta\,  \sqrt{ \langle\bar u^2\rangle_\tau},\quad
         k=\frac{2\pi}{\lambda}.
\end{equation}
Subscript $\tau$ means that we are interested in {\em variation} of
temperature during observation time $\tau$.

The total variation of TD temperature fluctuations is equal to
$$
\langle\bar u^2\rangle = \frac{k_B T^2}{\rho C \, \pi R_b^2l_c}
$$
where $k_B$ is the Boltzmann constant, $\rho$ is density, and $C$ is the
specific heat capacity. Then the variation of temperature over
the small  time $\tau$ (adiabatic approximation) must be about
$ \langle\bar u^2\rangle_\tau \simeq \langle\bar
u^2\rangle \times \tau/\tau^* $ where $\tau^*=\rho C R_b^2/\kappa$ is thermal
relaxation time of our cylinder through lateral surface (base surfaces of
cylinder are thermo isolated), $\kappa$ is thermal
conductivity. Here we assume that
$\tau \ll \tau^*$. This result can be rewritten in form
$$ \langle\bar
u^2\rangle_\tau = \frac{k_B T^2}{\rho C \, \pi R_b^2l_c}\times
\left(\frac{r_T^2}{R_b^2}\right),\quad r_T =\sqrt \frac{\kappa \tau}{\rho C}
\ll R_b,\ l
$$
where $r_T$ is thermal diffusive length for the time  $\tau$.

Equating
$
\langle\bar u^2\rangle_\tau\simeq S_{\bar u}(\omega)\, \Delta \omega
$ we can obtain the estimate for spectral density $S_{\bar u}(\omega)$ of
averaged temperature putting $\omega\simeq \Delta \omega\simeq 1/\tau$.

The accurate expressions for these spectral densities $S_u(\omega)$
and $S_\varphi(\omega)$ of temperature
fluctuations  and  phase fluctuations correspondingly are the following
\begin{align}
\label{Su}
S_{\bar u}(\omega)&\simeq
	\frac{4 k_B T^2\kappa }{ (\rho C)^2 l_c }\,
	\frac{1}{\pi R_b^4\omega^2},\\
\label{Sphi}
S_{\varphi}(\omega)&\simeq
	\frac{4 \beta^2 k^2 l_c  k_B T^2\kappa }{ (\rho C)^2  }\,
	\frac{1}{\pi R_b^4\omega^2}
\end{align}
for adiabatic case, i.e. for
$\omega\gg \frac{\kappa R_b^2}{\rho C}$.

One can easy check that our estimate  differs from accurate
expression (\ref{Su}) for $S_{\bar u}(\omega)$ only by the factor of about unity.
In Appendix \ref{Athermo} we present derivation of general expression for
adiabatic and non-adiabatic cases.

The formulae (\ref{Su}, \ref{Sphi}) allows to recalculate thermo-refractive
fluctuations into the fluctuations of dimensionless metric $h$ (which usually
describes the  sensitivity of laser gravitational antennae):
$
h=\varphi/(kL),
$
where $L$ is cavity length. It is useful to estimate its spectral
density $S_h(\omega)$ for
parameters of laser gravitational antenna (advanced LIGO)
presented in Appendix \ref{param} and $l_c=10$~cm:
\begin{align}
\label{Sh}
\sqrt{S_{h}(\omega)} & \simeq  0.5\times 10^{-24}\ \mbox{Hz}^{-1/2}.
\end{align}
It is about $4$ times smaller than the sensitivity of Standard Quantum Limit
(\ref{ShSQL}) which is planned to achieve in Advanced LIGO \cite{ligo2}.
Important that this thermo-refractive noise  rapidly decreases with increase
of beam radius: $\sqrt{S_{h}(\omega)}\sim R_b^{-2}$. Thus the using
so called "mesa-shaped" beams \cite{our,our2} (having flat distribution in the
center and fall to zero more quickly than Gaussian distribution at the edges)
with larger radius will allow to decrease thermo-refractive noise by several
times. For example using the $45\times 45$~cm$^2$ foot of CR and $R_b=10$~cm
with mesa shape distribution of the intensity of light in the beam one may
expect the gain of sensitivity for $\sqrt{S_{h}(\omega)}$ approximately one
order better than $\sqrt{S_{h}^{SQL}(\omega)}$.

\section{Conclusion}

The presented analysis of FP cavity with two CR (instead
of mirrors) has to be regarded as an example of cavity in which thermoelastic
noise in the coating may be substantially decreased and which permits to
circumvent substantially the SQL of sensitivity and also to have $(1-R)\approx
10^{-5}$.

We have shown that CR cavity is considerably more stable
than cavity with spherical mirrors relative to tilt and displacement
distortion. The distortion due to expose angle of CR is not so
small but it depends only on manufacturing procedure, and there is a hope
to decrease it due to improvement of manufacturing culture.

There does exist another argument in favor of using CR in FP resonators. The
very recent measurements performed by LIGO collaborators from University of
Glasgow, Stanford University, Iowa State University, Syracuse University and
LIGO Lab have shown that multi-layer coating also decreases the quality
factors of mirrors internal modes \cite{collab}. This effect may
substantially increase the Brownian component of the noise in the mirror
itself and thus decrease the sensitivity of LIGO antennae (see details on
Brownian moise in coating in \cite{etal1,etal2,etal3,etal4,etal5,etal6}).

At the same time it is likely that there does exist other version of cavity
free
from thermoelastic noise in coating, probably more easy to implement which
evidently deserves similar in-depth analysis. One of the "candidates" is a
cavity with unusual reflective coating: each layer of it has to have the same
small value of thermal expansion as fused silica. Unfortunately the technology
which may provide it is not yet invented.

In the above analysis  we have limited ourselves to the calculations of the
cavity properties itselves and did not discuss the coupling of cavity with
pumping laser and readout system. This analysis has to be done especially
because the readout system may be an intra-cavity one \cite{f1} and because
special attention has to be paid to the possible specific deformation of the
main mode distribution in FP cavity with CR. One of several
possible ways to realize the coupling of the mode with pumping source may be
based on the existence of the evanescent optical field "outside" the surface
of the facets. In this case it will be evidently necessary to use very thin
dielectric grating on the surface of the facet.

We have not also discuss the polarization characteristics of CR.  For example
it is known that the phase shift of  wave reflected from plane surface
depends on polarization \cite{ll}) and hence the FP cavity with 2-facets CR
will have slightly different eigen frequencies for waves with polarization
along and perpendicular to edge of CR. The additional problem to be analyzed
is the polarization characteristics of 3-facets CR.

It seems that the CR cavity for Advanced LIGO have to be used not for modes
with Gaussian distribution of power over the cross section but for so called
"mesa-shaped" \cite{our,our2} mode with flat distribution in the center and
fall to zero more quickly than Gaussian distribution at the edges.  For
"mesa-shaped" modes the profile of "lenses" ($h_1$ and $h_2$ in \ref{G0a}) on
the foot of the corner reflector must have special dependence on radius
calculated in \cite{our,our2}.

It is worth noting that the discussed in this paper features of CR cavity
may be useful not only for the gravitational wave
antennae but also in other high resolution spectroscopic experiments
where the low level of optical eigenmode fluctuations is important.

\section*{Acknowledgments}
We are very grateful to A.~S.~Ilinsky and
F.~Ya.~Khalili for very fruitful notes and
discussions. This work was supported by LIGO team from Caltech and
in part by NSF and Caltech grant PHY0098715, by
Russian Ministry of Industry and Science contracts \#\, 40.02.1.1.1.1137 and
\# \,40.700.12.0086, and
by Russian Foundation of Basic Researches grant \#03-02-16975-a.

\appendix

\section{Tilt of corner reflection}\label{Atilt}

The tilt of  one reflector through angle $\theta$ (see fig.\ref{tilt}a) can be
considered as displacement of lens  on distance $\delta x\simeq l\, \theta$
(see fig.\ref{tilt}b) perpendicular to axis. Then the set of integral equations
(\ref{ideal1}, \ref{ideal}) for eigen mode have the same form with
replacing kernel $G_0\, \to\, G_1$:
\begin{align}
G_1&\simeq  G_0
    \left(1-\frac{ix_1\,\delta x\, }{r_h^2}\right).
\end{align}

The perturbed main modes
$\breve\Phi_1^{00}(x_1,y_1)$,  $\breve\Phi_2^{00}(x_2,y_2)$ we find as
expansion into series
over eigen functions of resonator with perfectly positioned reflectors:
\begin{align}
\breve\Phi_1^{00} (x_1,y_1) &=
	\phi_0(y_1)  \sum_{m}(-1)^m\alpha_{m}\phi_{m}(x_1)\,  ,\\
\breve\Phi_2^{00} (x_2,y_2) &=
	\phi_0(y_2)  \sum_{m}\beta_{m}\phi_{m}(x_2)\,  .
\end{align}
After substitution these expansions into set (\ref{ideal1}, \ref{ideal})
replacing kernel $G_0\, \to\, G_1$ we obtain:
\begin{align}
\label{eigen1t}
&\sum_{m} \lambda_{m}(-1)^m\alpha_m\phi_m(x_2)-\\
&\quad -	e^{ikL}
	\int G_0(\vec r_1,\vec r_2)\,
	\left(\frac{ix_1\,\delta x}{r_h^2}\right)\times\nonumber\\
&\qquad \times        \sum_{m} (-1)^m \alpha_{m} \phi_{m}(x_1)\,dx_1=
	\nonumber\\
&=    \Lambda\, \sum_{m}(-1)^m \beta_{m}\phi_{m}(x_2)\nonumber,
\end{align}
\begin{align}
\label{eigen2t}
&\sum_{m} \lambda_{m}\beta_m\phi_m(x_1)-	e^{ikL}
	\int G_0(\vec r_1,\vec r_2)\times\\
& \qquad \times
	\left(\frac{ix_1\,\delta x}{r_h^2}\right)
        \sum_{m}  \beta_{m} \phi_{m}(x_2)\,dx_2=\nonumber\\
&=    \Lambda\, \sum_{m} \alpha_{m}\phi_{m}(x_1)\,         \nonumber
\end{align}
Here $\Lambda$ is eigen value of perturbed main mode.

After multiplying equation (\ref{eigen1t}) by
$\phi_{m_0}(x_2)$ and integrating we obtain:
\begin{align}
\label{eigen1ta}
\Lambda\,  \beta_{m_0}&=
	\lambda_{m_0,0}\alpha_{m_0} + I_{m_0}\, ,\\
I_{m_0,0} &=\frac{i\,\delta x\, r_L}{\sqrt 2\,  r_h^2}\, \lambda_{m_0,0}
	\times\nonumber\\
&\qquad \times	\left(\alpha_{m_0-1}  \sqrt {m_0} +
	\alpha_{m_0+1}  \sqrt {m_0+1}	\right)\nonumber
\end{align}
After multiplying equation (\ref{eigen2t}) by
$\phi_{m_0}(x_1)$ and integrating we obtain:
\begin{align}
\label{eigen1tb}
\Lambda\, {m_0} \alpha_{m_0}&=
	\lambda_{m_0,0}\beta_{m_0} + J_{m_0,0}\,    ,\\
J_{m_0,0} &=  \frac{i\delta x\,r_L}{\sqrt 2 \, r_h^2}
	\left(\lambda_{m_0-1}\, \beta_{m_0-1}\,\sqrt{m_0}+
	\frac{}{}\right.\nonumber\\
&\qquad \left. \frac{}{}
	\lambda_{m_0+1}\, \beta_{m_0+1}\,\sqrt{m_0+1}\right)\nonumber
\end{align}

Now we can rewrite equations (\ref{eigen1ta},\ref{eigen1tb}) for different
$m_0$ taking in mind that $\alpha_0,\ \beta_0 \simeq 1$, $\alpha_1\simeq\
\beta_1 = {\cal O}(\delta x)$, $\alpha_2\simeq \beta_2 = {\cal O}(\delta x^2),\ \dots$:
\begin{align}
m_0=0, &\quad \lambda_0\, \alpha_0 +
	\frac{i\delta x\,r_L}{\sqrt 2 \, r_h^2}\,\lambda_0\alpha_1=
	\Lambda\,\beta_0,\\
& \Lambda \alpha_0= \lambda_0\beta_0+
	\frac{i\delta x\,r_L}{\sqrt 2 \, r_h^2}\, \lambda_1\beta_1,\\
\Rightarrow & \alpha_0\simeq \beta_0\simeq 1,\quad
	\Lambda =\lambda_0 +{\cal O}(\delta x ^2),
\end{align}
\begin{align}
m_0=1,&\quad  \lambda_1 \alpha_1 -\Lambda \beta_1 \simeq
	\frac{-i\delta x\,r_L}{\sqrt 2 \, r_h^2}\,\lambda_1\, ,\\
&	-\Lambda \alpha_1 +\lambda_1\, \beta_1\simeq
	\frac{-i\delta x\,r_L}{\sqrt 2 \, r_h^2}\,\lambda_0\, ,\\
\Rightarrow & \alpha_1 \simeq
	\frac{-i\delta x\,r_L\, \big(\lambda_1^2+\lambda_0^2\big)}
	{\sqrt 2 \, r_h^2\big(\lambda_1^2-\lambda_0^2\big)},\\
\Rightarrow &	\beta_1 \simeq
	\frac{-i\delta x\,r_L\, \lambda_1}
	{\sqrt 2 \, r_h^2\big(\lambda_1-\lambda_0\big)},
\end{align}
\begin{align}
m_0=2,&\quad 	 \lambda_2 \alpha_2 -\Lambda \beta_2 \simeq
	\frac{-i\delta x\,r_L}{\sqrt 2 \, r_h^2}\,
	\lambda_2\,\sqrt 2\, \alpha_1 ,\\
&	-\Lambda \alpha_2 +\lambda_2\, \beta_2\simeq
	\frac{-i\delta x\,r_L}{\sqrt 2 \, r_h^2}\,
	\sqrt 2\, \lambda_1\,\beta_1 ,\\
\Rightarrow &	\alpha_2\simeq \frac{-i\delta x\,r_L}{\, r_h^2}
	\frac{\Big(\lambda_2^2 \,\alpha_1 +\lambda_1 \lambda_0\, \beta_1\Big)}
	{\Big(\lambda_2^2-\lambda_0^2\Big)},\\
\Rightarrow &
	\beta_2 \simeq \frac{-i\delta x\,r_L}{\, r_h^2}\,
	\frac{\lambda_2\Big(\lambda_0 \,\alpha_1 +\lambda_1 \, \beta_1\Big)}
	{\Big(\lambda_2^2-\lambda_0^2\Big)},
\end{align}
Rewriting values $\alpha_1$ and $\beta_1$ using $g$-parameter (\ref{g1} --
\ref{g3}) one can obtain
formulas (\ref{atilt1} -- \ref{btilt1b}).

\section{Displacement of CR}\label{Adisp}

Let the right corner is displaced by value $\delta x$ in transversal
direction (see fig.~\ref{disp2}).
Then the integral equations for perturbed eigen mode  is the following
\begin{align}
\label{eq1}
e^{ikL}\int G_1(x_1, y_1, x_2, y_2)\,\breve\Phi_1(x_1, y_1)\, dx_1\, dy_1 &=\\
    =\Lambda\, \breve\Phi_2(\delta x-x_2, -y_2), \\
\label{eq2}
e^{ikL}\int G_1(x_1, y_1, x_2, y_2)\,\breve\Phi_2(x_2,y_2)\, dx_2\, dy_2 &=\\
    =\Lambda\,
        \breve\Phi_1(-\delta x-x_1, -y_1),
\end{align}
\begin{align}
\label{eq3}
\breve\Phi_1(-\delta x-x_1, -y_1) &\simeq    \breve\Phi_1(-x_1,-y_1) -\\
&-    \delta x\, \partial_{x_1}\breve\Phi_1(-x_1-y_1),\nonumber\\
\breve\Phi_2(\delta x-x_2, -y_2) &\simeq      \breve\Phi_2(-x_2,-y_2) +\\
&+    \delta x\, \partial_{x_2}\breve\Phi_2(-x_2-y_2),   \nonumber\\
\label{G1}
G_1(x_1, y_1, x_2, y_2) &\simeq
	G_0(x_1, y_1, x_2, y_2)\times\\
&\times \Big(1-i\big[\delta h_1+\delta h_2\big]\Big) ,\nonumber\\
\delta h_1 &\simeq \frac { x_1 \delta x}{2r_h^2},\quad
    \delta h_2 \simeq \frac {- x_2 \delta x}{2r_h^2}.
\end{align}
We find perturbed main mode distributions
$\breve\Phi_1^{00}(x_1,y_1), \ \breve\Phi_2^{00}(x_2,y_2)$ as expansion
into series over eigen functions of resonator with perfectly positioned
reflectors:
\begin{align}
\Phi_1^{00} (x_1,y_1) &= \phi_0(y_1)  \sum_{m}(-1)^m\alpha_{m}\phi_{m}(x_1)\,  \\
\Phi_2^{00} (x_2,y_2) &=
	\phi_0(y_2)  \sum_{m}\beta_{m}\phi_{m}(x_2) .
\end{align}
After substitution these expansions into (\ref{eq1} - \ref{G1}) we obtain:
\begin{align}
&\sum_{m} \lambda_{m,0}(-1)^m\alpha_m\phi_m(x_2)-	e^{ikL}
	\int G_0(\vec r_1,\vec r_2)\times\nonumber\\
&\qquad \times	\left(\frac{i(x_1-x_2)\delta x}{2r_h^2}\right)
        \sum_{m} (-1)^m \alpha_{m} \phi_{m}(x_1)\,dx_1=\nonumber\\
\label{eigen1}
=&    \Lambda\, \sum_{m}(-1)^m \beta_{m}\phi_{m}(x_2)+\\
&\quad    +\Lambda\,\delta x\, \sum_{m}(-1)^m\beta_{m}
    \left(\partial_{x_2}\phi_{m}(x_2)\right),\nonumber
\end{align}
\begin{align}
\label{eigen2}
&\sum_{m} \lambda_{m,0}\beta_m\phi_m(x_1)-   e^{ikL}
	\int G_0(\vec r_1,\vec r_2)\times\\
&\qquad \times	\left(\frac{i(x_1-x_2)\delta x}{2r_h^2}\right)
        \sum_{m}  \beta_{m} \phi_{m}(x_2)\,dx_2=\nonumber\\
=&    \Lambda\, \sum_{m} \alpha_{m}\phi_{m}(x_1)
	-\Lambda\,\delta x\, \sum_{m}\alpha_{m}
    \left(\partial_{x_1}\phi_{m}(x_1)\right),\nonumber
\end{align}

After multiplying equation (\ref{eigen1}) by
$\phi_{m_0}(x_2)$ and integrating we obtain:
\begin{align}
\label{eigen1a}
&\lambda_{m_0,0}\alpha_{m_0,0} + I_{m_0,0}=
	\Lambda\, \beta_{m_0,0}+ J_{p_0,0},\\
I_{m_0,0} &=\frac{i\delta x\, r_l}{2\sqrt 2\,  r_h^2}\left(
	\big[\lambda_{m_0,0}-\lambda_{m_0-1,0}\big]\,
		\sqrt{m_0} \alpha_{m_0-1}+\frac{}{}\right.\nonumber\\
&\qquad \left.+	\big[\lambda_{m_0,0}-\lambda_{m_0+1,0}\big]\,
		\sqrt{m_0+1} \alpha_{m_0+1}\right),\nonumber\\
J_{m_0,0}&=-\frac{\Lambda\,\delta x}{r_L}\left(
	\beta_{m_0+1}\,\sqrt {\frac {m_0+1}{2}}-
	\beta_{m_0-1}\,\sqrt {\frac {m_0}{ 2}}\right).\nonumber
\end{align}
After multiplying equation (\ref{eigen2}) by $\phi_{m_0}(x_1)$ and
integrating we obtain:
\begin{align}
\label{eigen2a}
&\lambda_{m_0,0}\beta_{m_0,0} + I_{m_0,0}'=
    \Lambda\, \alpha_{m_0,0} + J_{p_0,0}',\\
I_{m_0,0}' &= \frac{i\delta x\, r_l}{2\sqrt 2\,  r_h^2}\left(
	\big[\lambda_{m_0,0}-\lambda_{m_0-1,0}\big]\,
		\sqrt{m_0} \alpha_{m_0-1}+\right.\frac{}{}\nonumber\\
&\qquad +\left.	\big[\lambda_{m_0,0}-\lambda_{m_0+1,0}\big]\,
		\sqrt{m_0+1} \alpha_{m_0+1}\right),\nonumber\\
J_{m_0,0}'&= \frac{-\Lambda\,\delta x}{\sqrt 2\, r_L}\left(
	\alpha_{m_0+1}\,\sqrt {m_0+1}-
	\alpha_{m_0-1}\,\sqrt {m_0}\right).\nonumber
\end{align}
Here we use the rule: coefficients $\alpha_m=0$ and $\beta_m=0$ if
$m< 0$.

Substituting $I_{m_0,0}$ and $J_{m_0,0}$ into
	(\ref{eigen1a}) and substituting
	$I_{m_0,0}'$ and $J_{m_0,0}'$ into (\ref{eigen2a}) we obtain two
equations. They may be transformed from one to other by replacement
$\alpha_m\, \to \, \beta_m$ and vice versa $\beta_m\, \to \, \alpha_m$.
So assuming that $\alpha_m = \beta_m$  we can solve only one equation:
\begin{align}
\label{equab}
&\lambda_{m_0,0}\alpha_{m_0} +
	\frac{i\delta x\, r_L}{2\sqrt 2\,  r_h^2}\times\\
&\quad\times\left(	\big[\lambda_{m_0,0}-\lambda_{m_0-1,0}\big]\,
		\sqrt{m_0} \alpha_{m_0-1}+\frac{}{}\right.\nonumber\\
&\qquad +\left.\frac{}{}	\big[\lambda_{m_0,0}-\lambda_{m_0+1,0}\big]\,
		\sqrt{m_0+1} \alpha_{m_0+1}\right)\nonumber\\
=&    \Lambda\, \alpha_{m_0} -
	\frac{\Lambda\,\delta x}{\sqrt 2\, r_L}\left(
	\alpha_{m_0+1}\,\sqrt {m_0+1}-
	\alpha_{m_0-1}\,\sqrt {m_0}\right),\nonumber
\end{align}

We assume that $\lambda_{0,0}=1$, $\Lambda=\lambda_{0,0}+\Delta=1+\Delta$,
$\alpha_0\simeq 1$, $\alpha_1\sim \delta x$, $\alpha_2 \sim \delta x^2, \ \dots$.
Putting $m_0=0$ in (\ref{equab}) we obtain $\Delta\sim \delta x^2$.
And putting  $m_0=1$ in (\ref{equab}) we  find
\begin{align}
\alpha_1 &\simeq -\alpha_0\, \frac{\delta x}{\sqrt 2\,  r_L}
	\left(
	\frac{ir_L^2}{2r_h^2}+\frac{1}{1-\lambda_{1,0}}
	\right)
\end{align}
Using (\ref{g1} -- \ref{g3}) one can rewrite  this formula in form
(\ref{alphadisp}).

\section{Expose perturbation of CR}\label{Aexpos}

For this case the  equations for eigen modes calculations are the following
in this section we do not mark by $\breve {}$ distribution function of
perturbed mode:
\begin{align}
\label{nonp1}
e^{ikL}\int G_0(\vec r_1, \vec r_2)\,\Phi_2(\vec r_2)\, d\vec r_2 &\simeq \\
	\simeq \Lambda \tilde \Phi_1(\vec r_1)\,
	\Big(1 -ib\gamma k|x_1|\Big),\nonumber\\
\label{nonp2}
e^{ikL}\int G_{0}(\vec r_1, \vec r_2)\, \Phi_1(\vec r_2)\, d\vec r_2 &=
    \simeq \Lambda\tilde \Phi_2(\vec r_2).\nonumber
\end{align}
We find the solutions as expansion
\begin{align}
\Phi_1 (x_1,y_1) &=  \phi_0(y_1)  \sum_{m}(-1)^m\alpha_{m}\phi_{m}(x_1)\\
\Phi_2 (x_2,y_2) &=
	\phi_0(y_2)  \sum_{m}\beta_{m}\phi_{m}(x_2)   .
\end{align}
Substituting them into equations (\ref{nonp1}, \ref{nonp2}) we obtain
\begin{align}
\label{eigenpc1}
\sum_{m} \lambda_{m,0}\beta_m\phi_m(x_1)
	&=    \\
 =\Lambda\,\sum_{m} \alpha_m\phi_m(x_1)
	\Big(1 -ib\gamma k|x_1|\Big),\nonumber\\
\label{eigenpc2}
\sum_{m} \lambda_{m,0}(-1)^m\alpha_m\phi_m(x_2)   &= 
    \Lambda\, \sum_{m} (-1)^m\beta_{m}\phi_{m}(x_2)
\end{align}
From last equation we obtain
$$
\beta_m =\frac{\lambda_{m,0}}{\Lambda}\, \alpha_m
$$
and substitute it into (\ref{eigenpc1}).
After multiplying obtained equation by $\phi_{m_0}(x_1)$
and integrating over $dx_1$ we get:
\begin{align}
\label{eigenpca3}
\lambda_{m,0}^2\alpha_{m_0}
	&=     \Lambda^2\, \alpha_{m_0}-
	ib\gamma \Lambda^2\, kr_L\sum_{m}\alpha_m\, F_{m_0,m},\\
F_{m_0,m} &= \int_{-\infty}^\infty |x|\, \phi_m(x)\,\phi_{m_0}(x)\, dx
\end{align}
We have tabulated coefficients $F_{m,m_0}$ --- result is presented in
Table~\ref{table}.

\begin{table}
\caption{Numerical values of coefficients $F_{m_0,m}$ for
	low indices}\label{table}\vspace{2mm}
\begin{tabular}{|l|llll|}
\hline
$\frac{}{}$	&$m= 0$& $m=2$ & $m=4$ & $m=6$ \\[2mm]
\hline
$m_0=0 \frac{}{}$
        & $1/\sqrt \pi$
	& $1/\sqrt {2\pi}$
	& $-1/(2\sqrt{6\pi})$
        & $1/(4\sqrt{5\pi})$ \\
\hline
\end{tabular}
\end{table}

Assuming that $\lambda_{0,0}=1$ and
$\alpha_0\simeq 1,\ \alpha_1,\ \alpha_2,\dots \ll 1$
we see that this system can be divided by two independent subsystems:
one for odd indices and another one --- for even indices. Odd indices can be
put zero and for even indices we have
\begin{align}
\Lambda^2&\simeq 1 +\frac{iL\gamma r_L}{b\sqrt \pi},\\
\alpha_2 &\simeq
	\frac{iL\gamma kr_L\,}{\sqrt{2\pi}\,b\big(1-e^{-8i\psi}\big)},\\
\alpha_4 &\simeq
	\frac{ iL\gamma kr_L\,}{2\sqrt{6\pi}\, b\,\big(1-e^{-16i\psi}\big)},
\end{align}
We see that all coefficients  $\alpha_2,\ \alpha_4,\dots  \sim \gamma$,
i.e. they have the same order over $\gamma$. However the convergence seems
to take place due to decreasing the coefficients $F_{m0,0}\sim 1/m_0^{5/4}$
with $m_0\to \infty$.

These expressions can be rewritten using $g$-parameter in form
(\ref{expos1}, \ref{expos2}).

\section{Diffractional losses on edge}\label{Adiff}

\begin{figure}
\psfrag{H}{$H$}
\psfrag{E}{$E$}
\psfrag{a}{$\alpha$}
\psfrag{t}{$\theta$}
\psfrag{x}{$x$}
\psfrag{y}{$y$}
\includegraphics[width=0.35\textwidth]{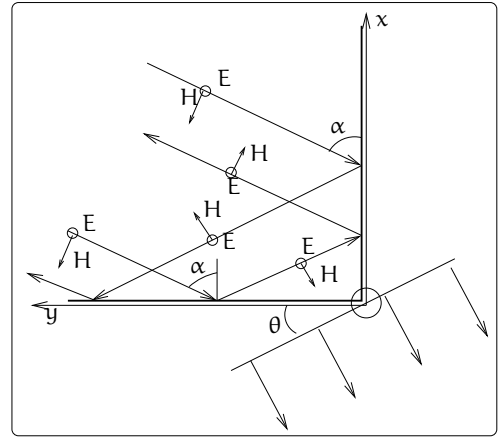}
\caption{Plane wave traveling and reflecting from dielectric corner
reflector with angle between facets $\pi/2$. The axis $z$ is directed upward
and perpendicular to plane of figure. Vector $E$ is directed along $z$-axis.
        }\label{cornerez}
\end{figure}

Here we write down the calculations to obtain estimate (\ref{funddiff}).
We consider the monochromatic plane wave polarized along $z$-axis
traveling and reflecting from
dielectric CR with angle between facets $\pi/2$  (see
fig.~\ref{cornerez}). Incident wave:
\begin{align*}
E_{z\, \rm inc}&=
        E_0 \, \exp\big(-i\omega t - ikn\sin\alpha \, x -
        ikn\cos\alpha \,y \big),\\
\vec H &= \big[\vec k\, \vec E\big], \quad\mu=1,
\end{align*}
Below I drop the multiplier $e^{-i\omega t}$.
Condition of internal reflection is fulfilled on both facets:
$ n \cos \alpha > 1,\qquad n \sin \alpha > 1.$
We assume that magnetic permittivity $\mu=1$ as in \cite{ll}
and hence $n^2=\epsilon$ ($\epsilon$ dielectric permittivity).

\paragraph*{Field inside CR.} We use Fresnel formulas for light wave reflection from plane boundary
between dielectric and vacuum using complex reflection coefficient $R_\bot$
\cite{ll} (for the case of complete internal reflection):
$$
R_\bot (\beta)=\frac{ncos\beta -i\sqrt{n^2sin^2\beta-1}}
        {ncos\beta +i\sqrt{n^2sin^2\beta-1}}
$$
where $\beta$ is incident angle.
The sum field after two reflections from both facets {\em inside} dielectric
is the following:
\begin{align}
\label{Ez}
E_{z}&= E_0 \left(
        e^{- i k_x x - ik_y y}+ R_\bot(\alpha) e^{+ ik_x x - ik_yy}+
         \frac{}{}\right.\\
&\quad + R_\bot(\pi/2-\alpha) e^{- ik_x x + ik_yy}+\nonumber\\
&\quad        + \left.\frac{}{}R_\bot(\pi/2-\alpha)R_\bot(\alpha)
                e^{+ ik_x x + ik_y y}
        \right) ,\nonumber\\
\label{kk}
k&=\frac{\omega}{c}, \quad k_x= kn\cos\alpha , \quad k_y=kn\sin\alpha
\end{align}
The first  terms in brackets describe the incident wave, the second and third
terms --- the waves reflected from planes ($y=0$) and ($x=0$) correspondingly,
the last term describes wave double reflected from both planes.

We simplify formula (\ref{Ez}) and calculate magnetic field:
\begin{align}
E_z &= 4E_0e^{-i(\delta_x+\delta_y)}
        \cos (k_x x-\delta_x)\, \cos (k_yy-\delta_y),\\
\tan \delta_x &=\frac{\sqrt{n^2\sin^2\alpha - 1}}{n\cos \alpha},\quad
        \tan \delta_y =\frac{\sqrt{n^2\cos^2\alpha -1}}{n\sin \alpha}
        \nonumber\\
\end{align}

\paragraph*{The field on outside surface of CR}
Now we can  write down the expressions for fields outside CR in
planes $x=0-\epsilon,\ y=0-\epsilon$ using boundary conditions ---  continuity
of tangent component of $\vec E$ and normal component $\mu \vec H$.
Then the expressions for fields are the following:
\begin{align*}
E_z^{x=0} &= 4E_0 e^{-i(\delta_x+\delta_y)}
        \cos  (\delta_x)\, \cos (k_yy-\delta_y),\\
 H_x^{x=0}& =
        4i\,nE_0\sin \alpha\, e^{-i(\delta_x+\delta_y)}
       \cos (\delta_x)\, \sin (k_yy-\delta_y),\,\\
H_y^{x=0} &= i4n E_0 \, \cos\alpha\, e^{-i(\delta_x+\delta_y)}
        \sin (\delta_x)\, \cos (k_yy-\delta_y),\\
E_z^{y=0} &= 4E_0e^{-i(\delta_x+\delta_y)}
        \cos (k_x x-\delta_x)\, \cos (\delta_y),\\
H_x^{y=0} &= -4in E_0 \, \sin\alpha \,e^{-i(\delta_x+\delta_y)}
       \cos (k_x x-\delta_x)\, \sin (\delta_y),\\
H_y^{y=0} &= -4inE_0 \, \cos\alpha\,e^{-i(\delta_x+\delta_y)}
        \sin (k_x x-\delta_x)\, \cos (\delta_y)\nonumber
\end{align*}

\begin{figure}
\psfrag{z}{$z$} \psfrag{f}{$\frac{\pi}{2}+\phi$}
\psfrag{t}{$\theta$} \psfrag{x}{$x$} \psfrag{y}{$y$} \psfrag{R}{$\vec r'$}
\includegraphics[width=0.3\textwidth]{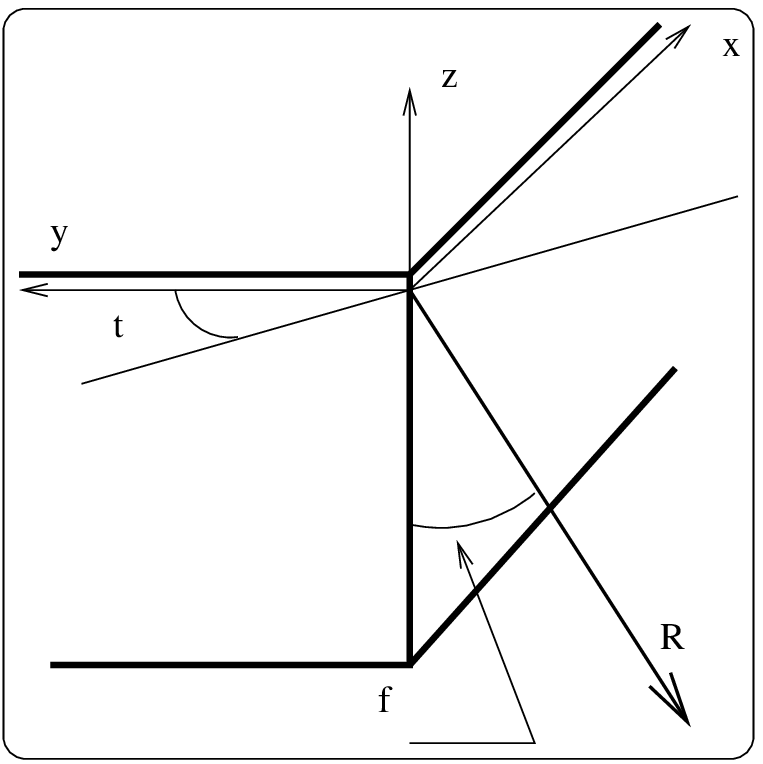}
\caption{ }\label{corner2}
\end{figure}

We take in mind that plane wave is limited by Gaussian multiplier
$$
b= \exp\Big(-a^2\big([x\sin\alpha -y\cos \alpha]^2 +z^2\big)\Big)
$$
with small parameter $a$, or more precisely: $a\ll k$.

\paragraph*{Radiation field.}
We can apply diffraction Green formula to  $E_z$ and calculate it in far wave zone in direction
characterized by angles $\phi,\ \theta$ and distance
$|\vec r -\vec r'|$ (see fig.{\ref{corner2}):
\begin{align}
E_z(\vec r')&= \int \left(E_z \partial_n G(R) -
	\frac{}{} G(R)\partial_n E_z\right)
        d\vec r,\\
G(R) &=\frac{e^{ikR}}{4\pi\,R},   \quad R=|\vec r'-\vec r|\gg \frac{1}{k} ,\\
\partial_n G(R)&\simeq -ikG(R)\times\frac{\vec n(\vec r'-\vec r)}{R}.
\end{align}
Here $\vec r'$ is radius-vector of observation point,  $\vec r$ is
radius-vector of  point on surface of integration, $d\vec r$ is element of
integration surface, $\vec n$ is normal to surface of integration.
The result of calculations is the following:
\begin{align}
E_z(R)&= \frac{iE_0\, e^{ikr'}\, e^{-i(\delta_x+\delta_y)}}{\pi  R}\times \\
&\qquad \times\underbrace{
	\frac{\sqrt \pi\, e^{-k^2\sin^2\phi/4a^2}}{a}
	}_{f(k\phi)\to \delta(k\phi)}\times \Big(I_x+I_y\Big),\nonumber\\
I_x&=   kb_x\int_0^\infty dx  \
        \cos (k_x x-\delta_x) e^{ik x cos\theta}\, e^{-a^2x^2\sin^2\alpha}
        \simeq
       \nonumber\\
&\simeq \frac{b_x\big(
        n\cos\alpha\, \sin \delta_x-i\cos\theta\, \cos\delta_x
        \big)} { n^2\cos^2\alpha - \cos^2\theta },\\
I_y&=  k b_y\int_0^\infty dy  \
        \cos (k_y y-\delta_y)\, e^{ik y \sin\theta}\, e^{-a^2y^2\cos^2\alpha}
        \simeq
	\nonumber\\
&\simeq\frac{b_y\big(
        n\sin\alpha\, \sin \delta_y-i\sin\theta\, \cos\delta_y
        \big)} { n^2\sin^2\alpha - \sin^2\theta }
\end{align}

Above we used auxiliary formulas:
\begin{align}
I_{c}(k) &=\int_0^\infty \cos kx\, e^{-a^2x^2}\, dx\ \Rightarrow \
	\pi\,\delta(k),\quad
	\mbox{if}\ a\ll k  ,\\
I_{s}(k) &=\int_0^\infty \sin kx\, e^{-a^2x^2}\, dx\
	\Rightarrow\
	\frac{1}{k},\quad
	\mbox{if}\ a\ll k  ,\nonumber
\end{align}

\paragraph*{Power of diffraction losses $W$} can be calculated as following
\begin{align}
W&=\int_{-\pi/2}^{\pi/2}\int_{-\pi/2}^{\pi}\,
        \frac{|E_z|^2c}{2\pi}\, R^2\, \cos\phi\, d\phi\, d\theta=
        \frac{ E_0^2 c}{2\pi^3}\times \frac{2\sqrt \pi}{ka}\times A,
        \nonumber\\
A&=\int_{-\pi/2}^{\pi} \big(I_x+I_y)^2\, d\theta
\end{align}
We have to compare the value $W$ with total power $W_0$ of incident wave
\begin{align}
W_0&= \frac{E_0^2c}{2a^2},\qquad
        (1-R)_d=\frac{W}{W_0} = \frac{2 a}{\pi^2\sqrt \pi\, k}\times A,
\end{align}
Replacing notations $a\to 1/\sqrt 2\,R_b$ and calculating numerically integral
$A\simeq 32.8$ for $n=1.45$ ($Si\, O_2$) and $\alpha=\pi/4$ we finally obtain
the estimate (\ref{funddiff}).

\section{TD temperature fluctuations in thermo-isolated
	layer}\label{Athermo}
We have thermal conductivity equation for temperature
$u(\vec r, t)$ in infinite layer with width $l$ ($0\le z\le l$) with
fluctuating force in right part \cite{bgv}
and the following boundary conditions:
\begin{align}
\label{TCl}
\frac {\partial u}{\partial t}-a^2 \Delta u&= F(\vec r,t),\\
\left.\frac{\partial u(\vec r, t)}{\partial z}\right|_{z=0,\, l}&=0,\\
\langle\, F(\vec r,t)F(\vec r',t')\,\rangle    & = -
	 \frac{2\kappa \, k_BT^2 }{ (\rho C)^2 }\,
	\Delta \delta (\vec r-\vec r')\, \delta (t-t'),
\end{align}
where $a^2=\kappa/\rho C$, $k_B$ is Boltzmann constant, $\delta$ is Dirac
delta function.

We find solution as series:
\begin{align*}
u(\vec r, t)&= \int\!\!\int\!\!\int_{-\infty}^{\infty}
	\frac{dk_x dk_y d\omega}{(2\pi)^3}
	\sum_{ n} u_{ n}(k_x, k_y,\omega)\times\\
&\quad	\times
	e^{i\omega t -ik_x x-ik_y y }\,\cos b_n z,\\
u_{ n}(k_x,k_y, \omega)&=
	\frac{F_{n}(k_x, k_y,\omega)}{i\omega+a^2(b_n^2+k_{\bot}^2)},\\
& b_n=\frac{\pi n}{l},\quad	k_{\bot}^2=k_x^2+k_y^2,\\
u_{ n}(k_x,k_y, \omega)&= \int\!\!\int\!\!\int_{-\infty}^{\infty}dx\,dy\,
	dt\,e^{-i\omega t +ik_x x +ik_y y}\times\\
&\times	\int_{0}^{l} dz \, \frac{2-\delta_{0,n}}{l}\, \cos b_n z\,
	 u(x,y,z,t),
\end{align*}
We find correlation functions of coefficients $F_{ n}(k_x, k_y,\omega)$:
\begin{align*}
F_{n,n_1} &=
	\langle\,F_{ n}(k_x,k_y, \omega)
	F^*_{ n_1}(k_{x_1},k_{y_1}, \omega_1)\,\rangle=\nonumber\\
&= {2(2\pi)^3k_B T^2 \kappa\over (\rho C)^2 }
	 \left(k_{\bot}^2+b_n^2\right)
	\frac{2-\delta_{0,n}}{l}\,\delta_{n,n_1}\,\times\\
&\quad \times
	\delta (k_{x}-k_{x_1})\,\delta(k_{y}-k_{y_1})\,\delta(\omega-\omega_1).
\end{align*}

We are interested in temperature $\bar u(t,x_0,y_0)$,
averaged over volume $V=\pi R_b^2 l$ along axis parallel to axis $z$ with
transversal coordinates $x_0$ and $y_0$, and also its correlation function
$\langle\, \bar u(t,0,0) \bar u(t+\tau),x_0,y_0\, \rangle $ with spectral
density $S_{\bar u}(\omega)$:
\begin{align}
\bar u(t,x_0,y_0)&=\frac{1}{\pi R_b^2 l}
	\int^{l}_{0} dz \int\!\!\int_{-\infty}^{\infty} dx\,dy\times
	\nonumber\\
&\times u(\vec r, t)
	e^{-\frac{(x-x_0)^2+(y-y_0)^2}{R_b^2} }=
	\nonumber\\
&=	\int^{l}_{0} \frac{dz}{l} \int\!\!\int_0^{\infty}
	\frac{dx\,dy}{\pi R_b^2}
	 e^{-\frac{(x-x_0)^2+(y-y_0)^2}{R_b^2} }\times \nonumber\\
&\times
	\int\!\!\int\!\!\int_{-\infty}^{\infty}
	\frac{dk_x dk_y d\omega}{(2\pi)^3}
	\sum_{ n} u_{ n}(k_x, k_y,\omega)\times \nonumber\\
&\times \cos b_n z\,
	e^{i\omega t -ik_x x-ik_y y }=\nonumber\\
&=	\int\!\!\int\!\!\int_{-\infty}^{\infty}
	\frac{dk_x dk_y d\omega}{(2\pi)^3}
	e^{-\frac{R_b^2k_{\bot}^2}{4} }\,\times \nonumber\\
&\times
	e^{i\omega t -ik_x x_0-ik_y y_0 }\,
	 u_{0}(k_x, k_y,\omega),
	\label{baru2}
\end{align}
\begin{align}
B_{\bar u}(\tau) &=
	\langle\, \bar u(t,0,0) \bar u(t+\tau,x_0,y_0)\, \rangle=\nonumber \\
&=
	\, \frac{2k_B T^2 \kappa}{ (\rho C)^2 }\,
	\frac{1}{l}
	\int\!\!\int\!\!\int_{-\infty}^{\infty}
	\frac{dk_x dk_y d\omega}{(2\pi)^3}\,\times \nonumber\\
&\times
	e^{-\frac{R_b^2k_{\bot}^2}{2} -ik_x x_0-ik_y y_0 }
	\frac{k_{\bot}^2e^{i\omega \tau}}{\omega^2+a^4 k_{\bot}^4}\,
	= 	\nonumber \\
&= \frac{k_B T^2 }{ (\rho C) }\, \frac{1}{\pi R_b^2 l(1+2a^2\tau/R_b^2)}\,
	e^{-\frac{x_0^2+y_0^2}{2(R_b^2+2a^2\tau)}},
\end{align}
\begin{align}
S_{\bar u}(\omega)&=2\int_{-\infty}^\infty d\tau \, e^{i\omega \tau}\,
	\langle\, \bar u(t,0,0) \bar u(t+\tau,0,0)\, \rangle=
	\nonumber\\
\label{STDTIb}
&=	\frac{4k_B T^2 \kappa}{ (\rho C)^2 l }
	\int_{0}^{\infty}
	\frac{k_\bot dk_\bot }{2\pi}\,
	e^{-\frac{R_b^2k_{\bot}^2}{2} }
	\frac{k_{\bot}^2}{\omega^2+a^4 k_{\bot}^4}\,
\end{align}
Making following substitutions
\begin{equation}
\label{subs}
	\xi=\frac{k_\bot^2R_b^2}{2}, \quad
		w=\frac{\omega R_b^2}{ 2\, a^2},\quad
	a^2=\frac{\kappa}{\rho C}
\end{equation}
one can express the spectral density using
exponential integrals:
\begin{align}
S_{\bar u}(\omega)&= 	\frac{k_B T^2 }{\pi \rho C l\, a^2 }
	\int_{0}^{\infty}
	\frac{ \xi\, d\xi}{w^2+\xi^2}\, e^{-\xi}=\nonumber\\
&=
	\frac{k_B T^2 }{2\pi \rho C l\, a^2 } \times \nonumber\\
&\times
	\Big(e^{iw}{\rm Ei}_1(iw)+ e^{-iw}{\rm Ei}_1(-iw) \Big)
	,\nonumber\\
{\rm Ei}_n(x) &= \int_1^\infty e^{-x\,t}\,\frac{dt}{t^n}
	\nonumber
\end{align}
For particular cases this formula can be simplified:
\begin{align}
\label{STDTInonad}
S_{\bar u}(\omega)|_{w\ll 1}&\simeq
	\frac{4 k_B T^2 \kappa}{ (\rho C)^2 l }\,
	\frac{1}{\pi R_b^4\omega^2},\\
\label{STDTIad}
S_{\bar u}(\omega)|_{w\gg 1}&\simeq
	\frac{4 k_B T^2 }{ \rho C\pi R_b^2 l }\times
	\frac{r_T^2}{R_b^2\omega}.
\end{align}
The formulas (\ref{STDTInonad} and \ref{STDTIad}) refer to non-adiabatic
and adiabatic cases correspondingly.

\section{Parameters}\label{param}
For our estimates we used the following parameters, material parameters
correspond to fused silica.
\begin{align*}
\omega&=2\pi\times 100\ \mbox{s}^{-1},\quad \lambda=1.064\ \mu\mbox{m},\quad
	T=300\ \mbox{K}, \\
b&= 2.3\ {\rm cm},\quad  R_b\simeq 6\ {\rm cm} \nonumber\\
m&=4\times 10^4\ \mbox{ g},\quad
	L=4\times 10^5\ \mbox{cm}; \label{parameter} \\
\alpha&=5.5\times  10^{-7}\ \mbox{K}^{-1}, \ \
	\kappa=1.4\times  10^5\ \frac{\mbox{erg}}{\mbox{cm s K}},\\
\rho&=2.2\ \frac{\mbox{g}}{\mbox{cm}^3}, \ \
	C=6.7\times  10^6\ \frac{\mbox{erg}}{\mbox{g K}},
	\nonumber \\
n&= 1.45,\quad \beta= \frac{dn}{dT} = 1.5\cdot 10^{-5}\ \mbox{K}^{-1}
\end{align*}


\end{document}